\documentclass[twocolumn,aps,prd,showpacs,nofootinbib]{revtex4}

\usepackage{amsmath}
\usepackage{amssymb}
\usepackage{graphicx}

\begin{document}

\title{Wormhole geometries supported by a nonminimal curvature-matter coupling}

\author{Nadiezhda Montelongo Garcia}
\email{nadiezhda@cosmo.fis.fc.ul.pt}\affiliation{Centro de Astronomia
e Astrof\'{\i}sica da Universidade de Lisboa, Campo Grande, Ed. C8
1749-016 Lisboa, Portugal}

\author{Francisco S.~N.~Lobo}
\email{flobo@cii.fc.ul.pt}\affiliation{Centro de Astronomia
e Astrof\'{\i}sica da Universidade de Lisboa, Campo Grande, Ed. C8
1749-016 Lisboa, Portugal}

\date{\today}

\begin{abstract}

Wormhole geometries in curvature-matter coupled modified gravity are explored, by considering an explicit nonminimal coupling between an arbitrary function of the scalar curvature, $R$, and the Lagrangian density of matter.
It is the effective stress-energy tensor containing the coupling between matter and the higher order curvature derivatives that is responsible for the null energy condition violation, and consequently for supporting the respective wormhole geometries. The general restrictions imposed by the null energy condition violation are presented in the presence of a nonminimal $R-$matter coupling. Furthermore, obtaining exact solutions to the gravitational field equations is extremely difficult due to the nonlinearity of the equations, although the problem is mathematically well-defined. Thus, we outline several approaches for finding wormhole solutions, and deduce an exact solution by considering a linear $R$ nonmiminal curvature-matter coupling and by considering an explicit monotonically decreasing function for the energy density. Although it is difficult to find exact solutions of matter threading the wormhole satisfying the energy conditions at the throat, an exact solution is found where the nonminimal coupling does indeed minimize the violation of the null energy condition of normal matter at the throat.

\end{abstract}


\maketitle

\section{Introduction}

Cosmology is said to be thriving in a golden age, where a central
theme is the perplexing fact that the Universe is undergoing an
accelerating expansion~\cite{expansion}. The late-time cosmic accelerated expansion is one of the most important and challenging current problems in cosmology, and represents a new imbalance in the governing gravitational
equations. The cause of this acceleration still remains an open and tantalizing question. Historically, physics has addressed such imbalances by
either identifying sources that were previously unaccounted for, or by altering the governing equations. The latter approach presents a very promising way to explain this major problem and assumes that at large scales Einstein's theory of General Relativity breaks down, where a more general action describes the gravitational field.
The Einstein field equation of General Relativity was first derived from an action principle by Hilbert, by adopting a linear function of the scalar curvature, $R$, in the gravitational Lagrangian density. However, there are no a priori reasons to restrict the gravitational Lagrangian to this form, and indeed several generalizations of the Einstein-Hilbert Lagrangian have been proposed, including ``quadratic Lagrangians'', involving second order curvature invariants such as $R^{2}$, $R_{\mu \nu }R^{\mu \nu }$, $R_{\alpha
\beta \mu \nu }R^{\alpha \beta \mu \nu }$, $\varepsilon ^{\alpha
\beta \mu \nu }R_{\alpha \beta \gamma \delta }R_{\mu \nu }^{\gamma
\delta }$, $C_{\alpha \beta \mu \nu }C^{_{\alpha \beta \mu \nu
}}$, etc \cite{early}. The physical motivations for these modifications of gravity were related to the possibility of a more realistic representation of the gravitational fields near curvature singularities and to create some first order approximation for the quantum theory of gravitational fields.

In this context, a more general modification of the Einstein-Hilbert gravitational Lagrangian density involving an arbitrary function of the scalar invariant, $f(R)$, was considered in \cite{Bu70}, and further developed in \cite{Ke81,Du83,BaOt83}. Recently, a renaissance of $f(R)$ modified theories of gravity has been verified in an attempt to explain the late-time accelerated expansion of the Universe \cite{Carroll:2003wy,Capozziello,NojOdint1,NojOdint2,Amendola}. We refer the reader to \cite{review} for reviews on this subject. It is also interesting to note that in the context of dark matter, the possibility that the galactic
dynamics of massive test particles may be understood without the
need for dark matter was also considered in the framework of
$f(R)$ gravity models \cite{darkmat}, and connections with MOND and the pioneer anomaly further explored by considering an explicit coupling of an arbitrary function of $R$ with the matter Lagrangian density
\cite{Bertolami:2007gv}. In the latter context, it was shown that a curvature-matter coupling induces a non-vanishing covariant derivative of the stress-energy, $\nabla_\mu T^{\mu\nu} \neq 0$.
Generally, the non-vanishing covariant divergence of the stress-energy tensor implies non-geodesic motion. The latter, due to the non-minimal couplings present in the model, imply the violation of the equivalence principle,
which is highly constrained by solar system experimental tests
\cite{Faraoni9,BPT06}. However, it has been recently reported, from
data of the Abell Cluster A586, that interaction of dark matter
and dark energy does imply the violation of the equivalence
principle \cite{BPL07}. Thus, it is possible to test these models with non-minimal couplings in the context of the violation of the equivalence
principle. It is also important to emphasize that the violation of the equivalence principle is also found as a low-energy feature of
some compactified version of higher-dimensional theories.

The equations of motion of test bodies for a theory with nonminimal coupling were also obtained by means of a multipole method \cite{Puetzfeld:2008xu}. This potentially leads to a deviation from geodesic motion, and consequently the appearance of
an extra force \cite{Bertolami:2007gv,Harko:2010zi}.
In this context, the perihelion precession of an elliptical planetary orbit in the presence of an extra force was obtained in a general form, and the magnitude of the extra gravitational effects was constrained in the case of a constant extra force by using Solar System observations \cite{Harko:2008qz}.
Implications, for instance, for stellar equilibrium have been studied in Ref. \cite{Bertolami:2007vu}. The equivalence with scalar-tensor theories with two scalar fields has been considered in Ref. \cite{Bertolami:2008im}, and a viability stability criterion was also analyzed in Ref. \cite{Faraoni:2007sn}. Analogous nonlinear gravitational couplings with a matter Lagrangian were also considered in the context of proposals to address the cosmic accelerated expansion \cite{Odintsov}, dark matter \cite{darkmatter}, in the analysis of the cosmological constant problem \cite{Lambda}, the curvature-matter coupling viability has been analyzed from the point of view of the energy conditions and of their stability under the Dolgov-Kawasaki criterion \cite{Bertolami:2009cd}, and matter density perturbations were analyzed to constrain the curvature-matter coupling theory from growth factor and weak lensing observations \cite{Nesseris:2008mq}.
In \cite{Harko:2010hw}, modified gravity with a nonminimal curvature-matter coupling in the Palatini formulation was considered, and in \cite{Harko:2010mv}, the $f(R)$ type gravity models was generalized by assuming that the gravitational Lagrangian is given by an arbitrary function of the Ricci scalar $R$ and of the matter Lagrangian $L_m$.
We refer the reader to Ref. \cite{Bertolami:2008zh} for a recent review on the nonminimal curvature-matter coupling.

In this paper, we explore the possibility that wormhole geometries be constructed by considering an explicit coupling between an arbitrary function of the scalar curvature, $R$, and the Lagrangian density of matter.
The coupling between the matter and the higher derivative curvature terms describes an exchange of energy and momentum between both, which is responsible for supporting the wormhole geometries.
The general restrictions imposed by the null energy condition violation are presented in the presence of a nonminimal $R-$matter coupling. Furthermore, obtaining exact solutions to the gravitational field equations is extremely difficult due to the nonlinearity of the equations, although the problem is mathematically well-defined. Thus, we outline several approaches for finding wormhole solutions, and deduce an exact solution by considering a linear $R$ nonmiminal curvature-matter coupling and by considering an explicit monotonically decreasing function for the energy density. Although it is difficult to find exact solutions of matter threading the wormhole satisfying the energy conditions at the throat, exact solutions are found where the nonminimal coupling does indeed minimize the violation of the null energy condition of normal matter at the throat.
The latter aspect has been given a particularly important treatment in the literature \cite{Visser:2003yf}, as it is a challenging feature of wormhole physics to minimize the amount of null energy condition violating matter threading the wormhole.

Note that in classical general relativity, wormholes are supported by exotic matter, which involves a stress-energy tensor that violates the null energy
condition (NEC) \cite{Morris:1988cz,Visser}. The NEC is given by $T_{\mu\nu}k^\mu k^\nu \geq 0$, where $k^\mu$ is {\it any} null vector. Several candidates have been proposed in the literature, amongst which we refer to solutions in higher dimensions, for instance in Einstein-Gauss-Bonnet theory \cite{EGB1,EGB2}, wormholes on the brane \cite{braneWH1}; solutions in Brans-Dicke theory \cite{Brans-Dicke}; wormholes constructed in $f(R)$ gravity; \cite{Lobo:2009ip}; wormhole solutions in semi-classical gravity (see Ref. \cite{Garattini:2007ff} and references therein); exact wormhole solutions using a more systematic geometric approach were found \cite{Boehmer:2007rm}; wormhole solutions and thin shells \cite{thinshells}; geometries supported by equations of state responsible for the cosmic acceleration \cite{phantomWH}; spherical wormholes were also formulated as an initial value problem with the throat serving as an initial value surface \cite{MontelongoGarcia:2009zz}; solutions in conformal Weyl gravity were found \cite{Weylgrav}, and thin accretion disk observational signatures were also explored \cite{Harko:2008vy}, etc (see Refs. \cite{Lemos:2003jb,Lobo:2007zb} for more details and \cite{Lobo:2007zb} for a recent review).

This paper is organized in the following manner: In Sec.
\ref{Sec:II}, we present a brief outline of the nonminimal curvature-matter coupling in $f(R)$ modified theories of gravity. In Sec. \ref{Sec:III}, we construct wormhole geometries supported by the nonminimal curvature-matter coupling, paying close attention to the energy conditions, and outlining different approaches in finding specific solutions. We further analyze in detail a particular solution by specifying the energy density threading the wormhole. We conclude in Sec. \ref{Sec:conclusion}.

\section{Nonminimal curvature-matter coupling in $f(R)$ gravity}
\label{Sec:II}

The action for a nonminimal curvature-matter coupling in modified theories of gravity, considered in this work \cite{Bertolami:2007gv}, is given  by
\begin{equation}
S=\int \left\{\frac{1}{2}f_1(R)+\left[1+\lambda f_2(R)\right]{\cal
L}_{m}\right\} \sqrt{-g}\;d^{4}x~,
\end{equation}
where $f_i(R)$ (with $i=1,2$) are arbitrary functions of the Ricci
scalar $R$ and ${\cal L}_{m}$ is the Lagrangian density corresponding to
matter. The coupling constant $\lambda$ characterizes the strength of the interaction between $f_2(R)$ and the matter Lagrangian.

Varying the action with respect to the metric $g_{\mu \nu }$
yields the field equations, given by
\begin{multline}
F_1(R)R_{\mu \nu }-\frac{1}{2}f_1(R)g_{\mu \nu }-\nabla_\mu
\nabla_\nu \,F_1(R)+g_{\mu\nu}\square F_1(R)\\
=-2\lambda F_2(R){\cal L}_m R_{\mu\nu}+2\lambda(\nabla_\mu
\nabla_\nu-g_{\mu\nu}\square){\cal L}_m F_2(R)\\
\hspace{1cm}+[1+\lambda f_2(R)]T_{\mu \nu }^{(m)}~,
\label{field}
\end{multline}
with the definition $F_i(R)=f'_i(R)$, where the prime represents
the derivative with respect to the scalar curvature. The
stress-energy tensor is defined as
\begin{equation}
T_{\mu \nu
}^{(m)}=-\frac{2}{\sqrt{-g}}\frac{\delta(\sqrt{-g}\,{\cal
L}_m)}{\delta(g^{\mu\nu})} ~.
\end{equation}

The contraction of the field equation (\ref{field}) yields the
trace equation
\begin{eqnarray}
F_1 R-2f_1+3\,\Box F_1 &=& -2\lambda\left(RF_2 {\cal L}_m
+ 3 \Box {\cal L}_m F_2 \right)
   \nonumber  \\
&&+\left( 1+\lambda f_2 \right) T^{(m)}
    \label{trace}
\end{eqnarray}
which shows that the Ricci scalar is a fully dynamical degree of
freedom.

Taking into account the covariant derivative of the field
Eqs. (\ref{field}), the Bianchi identities, $\nabla^\mu
G_{\mu\nu}=0$, and the identity
\begin{equation}
(\square\nabla_\nu -\nabla_\nu\square)F_i=R_{\mu\nu}\,\nabla^\mu F_i ~,
\end{equation}
one finally deduces the relationship
\begin{equation}
\nabla^\mu T_{\mu \nu }^{(m)}=\frac{\lambda F_2}{1+\lambda
f_2}\left[g_{\mu\nu}{\cal L}_m- T_{\mu \nu
}^{(m)}\right]\nabla^\mu R ~. \label{cons1}
\end{equation}
Equation (\ref{cons1}) shows that the coupling between the matter and the higher derivative curvature terms describes an exchange of energy and momentum between both \cite{Bertolami:2007gv}. It is interesting to note that analogous couplings arise after a conformal transformation in the context of scalar-tensor theories of gravity, and also in string theory. In the absence of the coupling, one verifies the conservation of the stress-energy tensor \cite{Koivisto,Bertolami:2007gv}, which can also be verified from the diffeomorphism invariance of the matter part of the action \cite{Faraoni,Carroll,Magnano}. Note that from Eq. (\ref{cons1}), the conservation of the stress-energy tensor is also verified if $f_2(R)$ is a constant or the matter Lagrangian is not an explicit function of the metric.

Equation (\ref{cons1}) provides non-geodesic motion governed by the following equation of motion for a fluid element: $dU^{\mu}/ds+\Gamma _{\alpha\beta}^{\mu}U^{\alpha}U^{\beta}=f^{\mu}$, where the extra force,
$f^{\mu}$, is given by
\begin{equation}
f^{\mu}=\frac{1}{\rho +p}\left[\frac{\lambda F_2}{1+\lambda
f_2}\left({\cal L}_m-p\right)\nabla_\nu R+\nabla_\nu p \right]
h^{\mu \nu }\,,
     \label{force}
\end{equation}
and $h_{\mu\nu}=g_{\mu\nu}+U_{\mu}U_{\nu}$ is the the projection operator \cite{Bertolami:2007gv}, and for simplicity, we have considered the stress-energy tensor for a perfect fluid
\begin{equation}
T_{\mu\nu}^{(m)}=\left(\rho +p\right)U_{\mu}U_{\nu}+pg_{\mu\nu}
\,,
   \label{perfectfluid}
\end{equation}
where $\rho$ is the energy density and $p$, the pressure,
respectively, and $U_{\mu }$ is the four-velocity.

However, a subtle issue needs to be pointed out. In a recent paper \cite{Sotiriou:2008it}, the authors argued that
a ``natural choice'' for the matter Lagrangian density for perfect
fluids is ${\cal L}_m=p$, based on Refs.
\cite{Schutz:1970my,Brown:1992kc}, where $p$ is the pressure. This
choice has a particularly interesting application in the analysis
of the curvature-matter coupling for perfect fluids, which imposes
the vanishing of the extra force \cite{Bertolami:2007gv}. Despite the fact that ${\cal L}_m=p$ does indeed reproduce the perfect fluid equation of state, it is not unique \cite{BLP}. Other choices include, for instance,
${\cal L}_m=-\rho$ \cite{Brown:1992kc,HawkingEllis,Faraoni:2009rk}, where $\rho$ is the energy density, or ${\cal L}_m=-na$, were $n$ is the particle number density, and $a$ is the physical free energy defined as $a=\rho/n-Ts$, with $T$ being the temperature and $s$ the entropy per particle (see Ref. \cite{BLP,Brown:1992kc} for details).

Therefore, no immediate conclusion may be extracted regarding the additional force imposed by the nonminimal coupling of curvature to matter, given the different available choices for the Lagrangian density. One may conjecture that there is a deeper principle or symmetry that provides a unique Lagrangian density for a perfect fluid \cite{BLP}. This has not been given due attention in the literature, as arbitrary gravitational field equations depending on the matter Lagrangian have not always been the object of close analysis. See Ref. \cite{BLP} for more details.

\section{Wormhole geometries supported by nonminimal curvature-matter couplings}\label{Sec:III}

\subsection{Spacetime metric and gravitational field equations}

Consider the following wormhole metric
\begin{equation}
ds^{2}=-e^{2\Phi(r)}dt^{2}+\frac{dr^{2}}{1-b(r)/r}+r^{2}(d\theta ^{2}+\sin
^{2}\theta d\phi ^{2})
  \label{WHmetric}
\end{equation}
where $\Phi(r)$ and $b(r)$ are arbitrary functions of the radial
coordinate, $r$, denoted as the redshift function, and the shape
function, respectively \cite{Morris:1988cz}. The radial coordinate
$r$ is non-monotonic in that it decreases from infinity to a
minimum value $r_0$, representing the location of the throat of
the wormhole, where $b(r_0)=r_0$, and then it increases from $r_0$
back to infinity.

A fundamental property of a wormhole is that a flaring out
condition of the throat, given by $(b-b^{\prime}r)/b^{2}>0$, is
imposed \cite{Morris:1988cz}, and at the throat
$b(r_{0})=r=r_{0}$, the condition $b^{\prime}(r_{0})<1$ is imposed
to have wormhole solutions. It is precisely these restrictions
that impose the NEC violation in classical general relativity.
Another condition that needs to be satisfied is $1-b(r)/r>0$. For
the wormhole to be traversable, one must demand that there are no
horizons present, which are identified as the surfaces with
$e^{2\Phi}\rightarrow0$, so that $\Phi(r)$ must be finite
everywhere. In the analysis outlined below, we consider that the
redshift function is constant, $\Phi'=0$, which simplifies the
calculations considerably, and provides interesting exact wormhole
solutions. If $\Phi'\neq 0$, the field equations become forth
order differential equations, and become intractable in the presence of a nonminimal curvature-matter coupling.

For simplicity, throughout this paper, we consider the specific case of $f_1(R)=R$, so that Eq. (\ref{field}) can be expressed in the following form
\begin{eqnarray}
&&R_{\mu \nu }-\frac{1}{2}R\,g_{\mu \nu }\equiv G_{\mu\nu}=
[1+\lambda f_2(R)]T_{\mu \nu }^{(m)} \nonumber \\
&&-2\lambda \left[ F_2(R){\cal L}_m R_{\mu\nu}-(\nabla_\mu
\nabla_\nu-g_{\mu\nu}\square){\cal L}_m F_2(R)\right]\,.
\label{field2}
\end{eqnarray}
Note that the curvature scalar, $R$, for the wormhole metric (\ref{WHmetric}), and considering $\Phi'=0$, is given by
\begin{eqnarray}
R&=& \frac{2b'}{r^2} \,.
    \label{Ricciscalar}
\end{eqnarray}

\subsection{Energy condition violations}

A fundamental point in wormhole physics is the energy condition
violations, as mentioned above. However, a subtle issue needs to
be pointed out in modified theories of gravity, where the
gravitational field equations differ from the classical
relativistic Einstein equations. Note that the energy conditions arise when one refers back to the Raychaudhuri equation for the expansion where a term
$R_{\mu\nu}k^\mu k^\nu$ appears, with $k^\mu$ any null vector. Imposing $R_{\mu\nu}k^\mu k^\nu\geq 0$ ensures that geodesic congruences
focus within a finite value of the parameter labelling points on
the geodesics. However, in general relativity, through the
Einstein field equation one can write the above condition in terms
of the stress-energy tensor given by $T_{\mu\nu}k^\mu k^\nu \ge
0$. In any other theory of gravity, one would require to know how
one can replace $R_{\mu\nu}$ using the corresponding field
equations. In particular, in a theory where we still have an Einstein-Hilbert term, the task of evaluating $R_{\mu\nu}k^\mu k^\nu$ is trivial. However, in $f(R)$ modified theories of gravity under consideration, things are not so straightforward.

For convenience Eq. (\ref{field2}) may be written as the following effective gravitational field equation
\begin{equation}
G_{\mu\nu}\equiv R_{\mu\nu}-\frac{1}{2}R\,g_{\mu\nu}= T^{{\rm
eff}}_{\mu\nu} \,,
    \label{field:eq2}
\end{equation}
where the effective stress-energy tensor is given by
\begin{eqnarray}
T^{{\rm eff}}_{\mu\nu}&=& [1+\lambda f_2(R)]T_{\mu \nu }^{(m)}
-2\lambda \big[ F_2(R){\cal L}_m R_{\mu\nu}
    \nonumber   \\
&&-(\nabla_\mu
\nabla_\nu-g_{\mu\nu}\square){\cal L}_m F_2(R)\big]\,.
    \label{efffield2}
\end{eqnarray}

In this context, the positivity condition, $R_{\mu\nu}k^\mu k^\nu \geq 0$, in the Raychaudhuri equation provides the following form for the null energy condition $T^{{\rm eff}}_{\mu\nu} k^\mu k^\nu\geq 0$, through the modified gravitational field equation (\ref{field}).

In principle, one may impose the condition $T^{(m)}_{\mu\nu} k^\mu k^\nu\ge 0$ for the normal matter threading the wormhole. This is useful as applying local Lorentz transformations it is possible to show that the above condition implies that the energy density is positive in all local frames of reference. This condition imposes interesting restrictions on the stress-energy tensor $T^{(m)}_{\mu\nu}$ threading the wormhole. From Eq. (\ref{efffield2}), the condition $T^{{\rm eff}}_{\mu\nu} k^\mu k^\nu< 0$, and considering that $1+\lambda f_2(R)>0$, imposes the restriction
\begin{eqnarray}
0\leq T_{\mu \nu }^{(m)}k^{\mu}k^{\nu}&<& \frac{2\lambda}{1+\lambda f_2(R)}
\big[ F_2(R){\cal L}_m R_{\mu\nu}k^{\mu}k^{\nu}
    \nonumber   \\
&&\hspace{-1cm}-k^{\mu}k^{\nu}(\nabla_\mu
\nabla_\nu-g_{\mu\nu}\square){\cal L}_m F_2(R)\big]\,.
    \label{effNEC1}
\end{eqnarray}
In order to have viable solutions, this inequality imposes the following additional restrictions
\begin{eqnarray}
F_2{\cal L}_m R_{\mu\nu}k^{\mu}k^{\nu}
    \gtrless-k^{\mu}k^{\nu}(\nabla_\mu
\nabla_\nu-g_{\mu\nu}\square){\cal L}_m F_2\,,
    \label{effNEC2}
\end{eqnarray}
for $\lambda \gtrless 0$. For the case of $1+\lambda f_2(R)<0$, then we simply arrive at the generic condition
\begin{eqnarray}
T_{\mu \nu }^{(m)}k^{\mu}k^{\nu}&>& \frac{2\lambda}{1+\lambda f_2(R)}
\big[ F_2(R){\cal L}_m R_{\mu\nu}k^{\mu}k^{\nu}
    \nonumber   \\
&&\hspace{-1cm}-k^{\mu}k^{\nu}(\nabla_\mu
\nabla_\nu-g_{\mu\nu}\square){\cal L}_m F_2(R)\big]\,.
    \label{effNEC1b}
\end{eqnarray}

However, in all the asymptotically solutions we have treated, and in particular, in the analysis presented below, it is extremely difficult to find exact asymptotically flat solutions in which $T^{(m)}_{\mu\nu} k^\mu k^\nu\ge 0$, due to the nonlinearity of the gravitational field equations. Nevertheless, we do present an exact solution where an increasing value of the nonminimal coupling parameter $\lambda$ does indeed minimize the NEC violation for the normal matter threading the wormhole throat.

\subsection{Specific case: $f_2(R)=R$}

Consider the specific case of $f_2(R)=R$, and we choose the Lagrangian form of ${\cal L}_m=-\rho(r)$, so that the field equation (\ref{field2}) reduces to
\begin{eqnarray}
G_{\mu\nu}= (1+\lambda R)T_{\mu \nu }^{(m)}+ 2\lambda \left[\rho R_{\mu\nu}-(\nabla_\mu \nabla_\nu-g_{\mu\nu}\square)\rho\right]\,.
\label{field3}
\end{eqnarray}

The stress-energy tensor for an anisotropic distribution of matter threading the wormhole is given by
\begin{equation}
T_{\mu\nu}=(\rho+p_t)U_\mu \, U_\nu+p_t\,
g_{\mu\nu}+(p_r-p_t)\chi_\mu \chi_\nu \,,
\end{equation}
where $U^\mu$ is the four-velocity, $\chi^\mu$ is the unit
spacelike vector in the radial direction, i.e.,
$\chi^\mu=\sqrt{1-b(r)/r}\,\delta^\mu{}_r$. $\rho(r)$ is the energy
density, $p_r(r)$ is the radial pressure measured in the direction
of $\chi^\mu$, and $p_t(r)$ is the transverse pressure measured in
the orthogonal direction to $\chi^\mu$.

Thus, the gravitational field equations are given by the following relationships
\begin{equation}
2\lambda \rho'' r(b-r)+\lambda \rho'(rb'+3b-4r)
    +\rho(r^2+2\lambda b')-b'=0\,,\label{field3ai}
\end{equation}
\begin{equation}
4\lambda r\rho'(b-r)+2\lambda \rho(b-b'r)
   -rp_r(r^2+2\lambda b')-b=0\,,
\label{field3aii}
\end{equation}
\begin{eqnarray}
4\lambda r^2\rho'' (b-r)+2\lambda r\rho'(rb'+b-2r)-2\lambda \rho(rb'+b)
    \nonumber  \\
-2rp_t\left(r^2+2\lambda b'\right)+b-rb'=0\,. \label{field3aiii}
\end{eqnarray}

The non-conservation of the stress-energy tensor, Eq. (\ref{cons1}), provides the following expression
\begin{eqnarray}
rp'_r(r^2+2\lambda b')+2\lambda rb''\left(\rho+p_r\right)
    \nonumber  \\
-4\lambda b'\left(\rho+p_t\right)+2r^2\left(p_r-p_t\right)=0\,.
\label{field4}
\end{eqnarray}

Relative to the energy conditions, considering a radial null vector, the violation of the NEC,
i.e., $T_{\mu\nu}^{{\rm eff}}\,k^\mu k^\nu < 0$, takes the
following form
\begin{eqnarray}
\rho^{{\rm eff}}+p_r^{{\rm eff}} &=&\frac{1}{r^2}\Bigg[-2\lambda r^2 \rho''\left(1-\frac{b}{r}\right)
    \nonumber  \\
&&\hspace{-2.0cm}+(\rho+p_r)\left(r^2+2\lambda b'\right)+\lambda \left(r\rho'+2\rho \right)\left(\frac{b'r-b}{r}\right)\Bigg],
     \label{NECeff}
\end{eqnarray}
where $\rho^{{\rm eff}}+p_r^{{\rm eff}}<0$. Using the
gravitational field equations, inequality (\ref{NECeff}) takes the
familiar form
\begin{equation}
\rho^{{\rm eff}}+p_r^{{\rm eff}}=\frac{b'r-b}{r^3}\,,
     \label{NECeff2}
\end{equation}
which is negative by taking into account the flaring out
condition, i.e., $(b'r-b)/b^2<0$, considered above.

At the throat, taking into account the finiteness of the factor $\rho''$ at the throat, one has the following relationship
\begin{eqnarray}
\left(\rho^{{\rm eff}}+p_r^{{\rm eff}}\right)\big|_{r_0}&=&
\frac{1}{r_0^2}\Big[(\rho_0+p_{r0})\left(r_0^2+2\lambda b'_0 \right)
    \nonumber   \\
&&-\lambda\left(r_0\rho'_0+2\rho_0 \right)(1-b'_0)\Big]\,,
     \label{NECeffb}
\end{eqnarray}
which implies the following general condition
\begin{eqnarray}
(\rho_0+p_{r0})\left(r_0^2+2\lambda b'_0 \right)
<\lambda\left(r_0\rho'_0+2\rho_0 \right)(1-b'_0)\,.
     \label{NECeffc}
\end{eqnarray}

One now has three equations, with four functions, namely, the field equations (\ref{field3ai})-(\ref{field3aiii}), with four unknown functions of $r$, i.e., $\rho(r)$, $p_r(r)$, $p_t(r)$ and $b(r)$. Obtaining explicit solutions to the Einstein field equations is extremely difficult due to the
nonlinearity of the equations, although the problem is mathematically well-defined. One may adopt different strategies to construct solutions with the properties and characteristics of wormholes. For instance, one may consider a plausible matter-field distribution, by considering convenient stress-energy tensor components, and through the field equations determine the metric field $b(r)$. In particular, one may specify an equation of state $p_r=p_r(\rho)$ or $p_t=p_t(\rho)$, thus closing the system.
The specific case of an isotropic pressure, $p_r=p_t$, also closes the system of Eqs. (\ref{field3ai})-(\ref{field3aiii}), but it is impossible to find an exact solution. Thus, in this work we only consider the anisotropic pressure case.

In principle, we can also adopt the approach in which a specific choice for a physically reasonable shape function $b(r)$ is provided and through Eq. (\ref{field3ai}), $\rho(r)$ is determined, thus consequently providing explicit expressions for the remaining stress-energy tensor components, through Eqs. (\ref{field3aii}) and (\ref{field3aiii}).

An alternative approach is to specify a physically reasonable form for the energy density, and thus solve Eq. (\ref{field3ai}) to find the shape function. The general solution of Eq. (\ref{field3ai}) for $b(r)$ is given by
\begin{eqnarray}
b(r)&=&\Bigg[\int \frac{r\exp[g(r)](-\rho r+2\lambda \rho''r+4\lambda \rho')}{\lambda(\rho'r+2\rho)-1}dr+C\Bigg]\times
     \nonumber  \\
     &&\times\exp[-g(r)]\,,\label{solb1}
\end{eqnarray}
where $C$ is a constant of integration and, for notational simplicity, the function $g(r)$ is defined as
\begin{equation}
g(r)=\lambda\int \frac{3\rho'+2\rho''r}{\lambda(\rho'r+2\rho)-1}dr \,.
\end{equation}
The radial pressure and the lateral pressure are consequently given by Eqs. (\ref{field3aii}) and (\ref{field3aiii}), and take the following form
\begin{equation}
p_r(r)=\left[4\lambda r\rho'(b-r)+2\lambda \rho(b-b'r)-b\right]/
\left[r(r^2+2\lambda b')\right]\,,
\label{prfield3aii}
\end{equation}
\begin{eqnarray}
p_t(r)&=&\big[4\lambda r^2\rho'' (b-r)+2\lambda r\rho'(rb'+b-2r)
   \nonumber  \\
&&\hspace{-0.5cm}-2\lambda \rho(rb'+b)+b-rb'\big]/\left[2r\left(r^2+2\lambda b'\right)\right]\,, \label{ptfield3aiii}
\end{eqnarray}
respectively.

\subsection{Specific solution: $\rho(r)=\rho_0(r_0/r)^\alpha$}


Adopting this latter approach, consider the specific energy density given by the following monotonically decreasing function
\begin{equation}
\rho(r)=\rho_0\left(\frac{r_0}{r}\right)^\alpha\,,\label{ex1}
\end{equation}
with $\alpha>0$, and $\rho_0>0$ is the energy density at the throat. Thus, taking into account Eq. (\ref{ex1}) and solving Eq. (\ref{solb1}), the shape function is consequently given  by
\begin{widetext}
\begin{eqnarray}
b(r)&=&\left(\frac{\rho_0r}{\alpha-3}\right)\left(\frac{r_0}{r}\right)^\alpha
\Bigg\{C
+2\alpha\lambda\left(\alpha-3\right)
{\rm hypergeom}\left(\left[\frac{\alpha-1}{\alpha},-\frac{1+\alpha}{\alpha-2} \right],\left[\frac{2\alpha-1}{\alpha} \right],-\rho_0 \lambda \left(\frac{r_0}{r}\right)^\alpha(\alpha-2)\right)
    \nonumber  \\
&&-r^2{\rm hypergeom}\left(\left[-\frac{\alpha+1}{\alpha-2},\frac{\alpha-3}{\alpha} \right],\left[\frac{2\alpha-3}{\alpha} \right],-\rho_0 \lambda \left(\frac{r_0}{r}\right)^\alpha(\alpha-2)\right) \Bigg\}
\left[\lambda\rho_0 \left(\frac{r_0}{r}\right)^\alpha \left(\alpha-2\right)+1\right]^{\frac{1-2\alpha}{\alpha-2}}
\,.\label{solb1b}
\end{eqnarray}

The constant of integration $C$ is deduced by considering $b(r_0)=r_0$. The spacetime is asymptotically flat, $b(r)/r\rightarrow 0$ for $r\rightarrow \infty$, if $\alpha \geq 3$.

In particular, imposing $\alpha=3$, then Eq. (\ref{solb1}) provides the following solution for the shape function
\begin{eqnarray}
\nonumber
b(r)&=&
\left[\frac{6}{7}\rho_{0} ^{5}\lambda ^{5}\frac{r_{0}^{15}}{r^{14}}-\frac{1}{12}\rho_{0} ^{5}\lambda^{4}
\frac{r_{0}^{15}}{r^{12}}+\frac{48}{11}\rho _{0}^{4}\lambda^{4}\frac{r_{0}^{12}}{r^{11}}
\right.-\frac{4}{9}\rho _{0}^{4}\lambda ^{3}\frac{r_{0}^{12}}{r^{9}}
+9\rho _{0}^{3}\lambda^{3}\frac{r_{0}^{9}}{r^{8}}-\rho _{0}^{3}\lambda^{2}\frac{r_{0}^{9}}{r^{6}}
    \nonumber  \\
&&+ \frac{48}{5}\lambda^{2}\rho _{0}^{2}\frac{r_{0}^{6}}{r^{5}}-\frac{4}{3}\rho _{0}^{2}\lambda
\frac{r_{0}^{6}}{r^{3}}+6\rho_{0}\lambda\frac{r_{0}^{3}}{r^{2}}
\left.+r_{0}^{3} \rho_{0}\ln(r)+C\right]
\left[1+\lambda\rho_{0}\left(\frac{r_{0}}{r}\right)^3\right]^{-5}\,.
\end{eqnarray}
The constant of integration is deduced by taking into account the condition $b(r_0)=r_0$, and is given by
 \begin{eqnarray}
C&=&r_0\Bigg\{\left[1+\lambda\rho_{0}\right]^{5}
    -\left[\frac{6}{7}\rho_{0} ^{5}\lambda ^{5}-\frac{1}{12}\rho_{0} ^{5}\lambda^{4}
r_{0}^{2}+\frac{48}{11}\rho _{0}^{4}\lambda^{4}
\right.-\frac{4}{9}\rho _{0}^{4}\lambda ^{3}r_{0}^{2}
+9\rho _{0}^{3}\lambda^{3}
   \nonumber  \\
&&-\rho _{0}^{3}\lambda^{2}r_{0}^{2}+ \frac{48}{5}\lambda^{2}\rho _{0}^{2}-\frac{4}{3}\rho _{0}^{2}\lambda
r_{0}^{2}+6\rho_{0}\lambda
 \left.+r_{0}^{2} \rho_{0}\ln(r_0)\right]\Big\}\,.
\end{eqnarray}
\end{widetext}

\begin{figure}[ht]
\includegraphics[width=7.0cm]{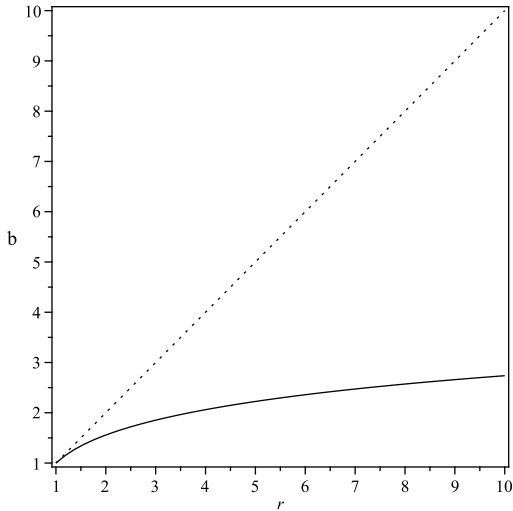}
\caption{The dashed curve represents the function $h(r)=r$, and the solid curve the shape function, $b(r)$. The shape function is plotted for the values $r_0=1$, $\rho_0=0.75$ and $\lambda=0.1$. The conditions $b/r\rightarrow 0$, as $r\rightarrow \infty$, and $b(r)<r$ are obeyed. See the text for details.}
\label{figure1}
\end{figure}

The shape function is plotted in Fig. \ref{figure1} for the values $r_0=1$, $\rho_0=0.75$ and $\lambda=0.1$. The fundamental wormhole conditions, namely, $b/r\rightarrow 0$, as $r\rightarrow \infty$, and $b(r)<r$ are obeyed. One may also deduce interesting restrictions from the derivative of the shape function evaluated at the throat, which is given by
\begin{equation}
b'_0=\frac{\rho_0(r_0^2+3\lambda)}{1+\rho_0\lambda} \,.
\end{equation}
Note that the condition at the throat $b'_0<1$ imposes the restriction $\lambda< (1-\rho_0 r_0^2)/(2\rho_0)$. For the values considered above, i.e., $r_0=1$, $\rho_0=0.75$ we have that the coupling parameter possesses the upper limit given by $\lambda<1/6$.

\begin{figure}[ht]
\includegraphics[width=7.0cm]{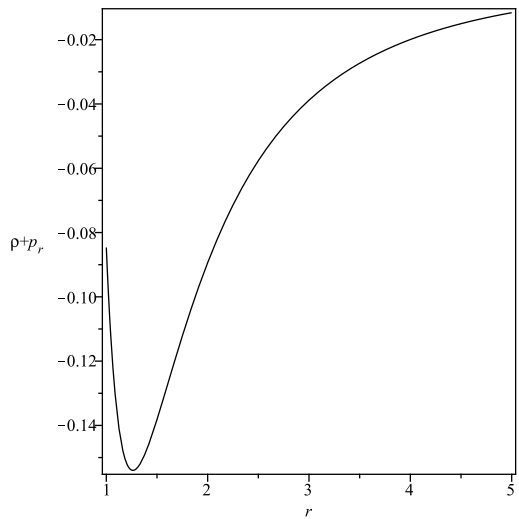}
\caption{The null energy condition (NEC), $\rho(r)+p_r(r)$, profile for the values $r_0=1$, $\rho_0=0.75$ and $\lambda=0.1$. See the text for details.}
\label{figure2}
\end{figure}

The NEC is violated for the normal matter threading the throat, as can be readily verified from Fig. \ref{figure2}.


It is interesting to compare this specific solution with the general relativistic case, where $\lambda=0$. Considering the the energy density profile given by Eq. (\ref{ex1}), the gravitational field equation (\ref{field3ai}) provides the following form function
\begin{equation}
b(r)=r_0+\rho_0 r_0^3 \ln\left(\frac{r_0}{r} \right).
\end{equation}
The geometry is asymptotically flat, i.e., $b/r\rightarrow 0$, as $r\rightarrow \infty$, and the condition $b(r)<r$ is also obeyed. Rather than reproduce the plot for $b(r)$, we refer that its profile is analogous to that of Fig. \ref{figure1}. The radial derivative of the shape function is given by $b'(r)=\rho_0 r_0^3/r$, so that the flaring out condition at the throat, $b'(r_0)<1$ imposes the restriction $\rho_0 r_0^2<1$.

The radial pressure and the transverse pressure are given by the following expressions
\begin{eqnarray}
p_r(r)&=&-\frac{r_0}{r^3}\left[1+\rho_0 r_0^2 \ln\left(\frac{r}{r_0} \right)\right]  \\
p_t(r)&=&\frac{r_0}{r^3}\left\{1+\rho_0 r_0^2 \left[\ln\left(\frac{r}{r_0} \right)-1\right]\right\}\,,
\end{eqnarray}
respectively.

\begin{figure}[ht]
\includegraphics[width=7.5cm]{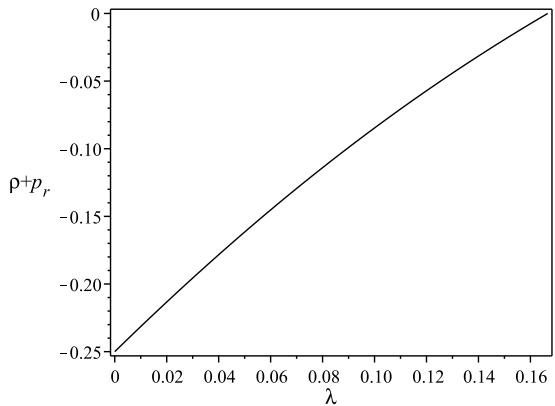}
\caption{The general relativistic deviation profile for the values $r_0=1$, $\rho_0=0.75$ and $\lambda=0.1$, considered at the throat. Note that it is possible to minimize the NEC violating matter for increasing values of $\lambda$. See the text for details.}
\label{figure3}
\end{figure}

A challenging endeavor in the literature has been to find wormhole solutions where normal matter minimizes the NEC violation at the throat. In this context, we note that the nonminimal coupling does in fact produce this effect by increasing the value of the parameter $\lambda$. To verify this, consider the NEC of matter threading the wormhole throat given by
\begin{eqnarray}
(\rho+p_r)|_{r_0}=\frac{\rho_0^2\lambda(2\lambda+r_0^2)
+\rho_0(\lambda+r_0^2)-1}{3\rho_0\lambda(2\lambda+r_0^2)+r_0^2}.
\label{NEClambda}
\end{eqnarray}
For concreteness, we take into account the specific values considered above given by $r_0=1$ and $\rho_0=0.75$, and from the condition $b'_0<1$, we verify that the positive nonminimal coupling parameter lies in the range $0<\lambda<1/6$. Equation (\ref{NEClambda}) is depicted in Fig. \ref{figure3}, where it is shown that is possible to minimize the NEC violating matter for increasing values of $\lambda$.

Thus, the example outlined in this section shows that one may in principle construct wormhole geometries supported by a nonminimal curvature-matter coupling, which is translated through an exchange of energy and momentum between both. Although it is difficult to find exact solutions of matter threading the wormhole satisfying the energy conditions at the throat, exact solutions were found where the nonminimal coupling minimized the violation of the null energy condition of normal matter at the throat.

\section{Conclusion}\label{Sec:conclusion}

Wormholes are hypothetical tunnels in spacetime, possibly through which
observers may freely traverse. However, it is important to emphasize that these solutions are primarily useful as ``gedanken-experiments'' and as a theoretician's probe of the foundations of general relativity. It is an important and intriguing challenge in wormhole physics to find a realistic matter source that will support these exotic spacetimes. In this work wormhole geometries in curvature-matter coupled modified gravity were explored, by considering an explicit nonminimal coupling between an arbitrary function of the scalar curvature, $R$, and the Lagrangian density of matter. The coupling between the matter and the higher derivative curvature terms describes an exchange of energy and momentum between both.

The general restrictions imposed by the null energy condition violation were presented in the presence of a nonminimal $R-$matter coupling. Furthermore, obtaining exact solutions to the gravitational field equations is extremely difficult due to the nonlinearity of the equations, although the problem is mathematically well-defined. Thus, we outlined several approaches for finding wormhole solutions, and deduced an exact solution by considering a linear $R$ nonmiminal curvature-matter coupling and by considering an explicit monotonically decreasing function for the energy density. Although it is difficult to find exact solutions of matter threading the wormhole satisfying the energy conditions at the throat, exact solutions were found where the nonminimal coupling minimized the violation of the null energy condition of normal matter at the throat.

In conclusion, it is well known that $f(R)$ modified theories of gravity are equivalent to a Brans-Dicke theory with a coupling parameter $w = 0$, and a specific potential related to the function $f(R)$ and its derivative. As a bridge to the work outlined in this paper, it would be interesting to analyze the equivalence between a suitable scalar theory and a model with a nonminimal curvature-matter coupling and its relationship to wormhole physics. Work along these lines in presently underway.

\acknowledgments

The authors acknowledge stimulating discussions with Orfeu Bertolami and Jorge P\'{a}ramos. NMG acknowledges a postdoctoral fellowship from CONACYT-Mexico. FSNL acknowledges financial support of the Funda\c{c}\~{a}o para a Ci\^{e}ncia e Tecnologia through the grants PTDC/FIS/102742/2008 and CERN/FP/109381/2009.

\end{document}